\documentclass[preprint,showpacs,preprintnumbers,amsmath,amssymb]{revtex4}

\usepackage[mathscr]{eucal}
\usepackage{amssymb}
\usepackage{amsmath}
\usepackage{amscd}
\usepackage{bm}
\usepackage[dvips]{graphicx}

\def\d3V#1#2{\frac{d^{#1}#2}{(2\pi)^{#1}}}

\def\Xa{\bm{X}_{i}}
\def\ua{u_i}
\def\va{v_i}

\def\vaq{v_i^3}
\def\uadt{\dot{u}_i}
\def\vadt{\dot{v}_i}
\def\Xat#1{\bm{X}_{i}(#1)}
\def\dotXat#1{\dot{\bm{X}}_{i}(#1)}

\begin{document}

\title{Averaging approach to phase coherence of uncoupled
    limit-cycle oscillators receiving common random impulses}%
\author{Kensuke Arai$^1$\footnote{E-mail: arai@ton.scphys.kyoto-u.ac.jp\\
    URL: http://www.ton.scphys.kyoto-u.ac.jp/nonlinear}
 and Hiroya Nakao$^{1,2}$ }
\affiliation{Department of Physics, Kyoto University, Kyoto 606-8502,
  Japan$^1$ \\
  Abteilung Physikalische Chemie, Fritz-Haber-Institut der
  Max-Planck-Gesellschaft, Faradayweg 4-6, 14195 Berlin, Germany$^2$ }
\date{\today}
             
\begin{abstract}
  Populations of uncoupled limit-cycle oscillators receiving
  common random impulses show various types of phase-coherent states,
    which are characterized by the distribution of phase
    differences between pairs of oscillators.
    We develop a theory to predict the stationary distribution of
    pairwise phase difference from the phase response curve, which
    quantitatively encapsulates the oscillator dynamics, via averaging
    of the Frobenius-Perron equation describing the impulse-driven
    oscillators.
    The validity of our theory is confirmed by direct numerical
    simulations using the FitzHugh-Nagumo neural oscillator receiving
    common Poisson impulses as an example.
\end{abstract}

\maketitle

\section{Introduction}

Coherence phenomena exhibited by dynamical units receiving correlated
drive signals has been the focus of much recent
research~\cite{Mainen-Sejnowski,Binder-Powers,Galan,Roy,Yoshida,Arai-Nakao,Yip-Uchida,Hudson,Goldobin-Pikovsky,Nagai-Nakao,Nakao-Arai,Teramae-Dan,Toral,Zhou-Kurths,Pakdaman,Kobayashi,Galan2}.
Experimentally, synchronization among dynamical units receiving common
fluctuating drive, or response reproducibility of a single unit
receiving identical fluctuating drive, has been shown in
neurons~\cite{Mainen-Sejnowski,Binder-Powers,Galan}, chaotic
lasers~\cite{Roy}, and electrical
oscillators~\cite{Yoshida,Yip-Uchida,Arai-Nakao}.
The slightly counterintuitive phenomenon of desynchronization or
anti-reliability via a common input has been seen in electrical
oscillators~\cite{Arai-Nakao}, electrochemical
oscillators~\cite{Hudson}, and light-sensitive circadian
cells~\cite{Kobayashi}.
Further, coexistence of multiple synchronized groups of dynamical
units have been observed in chaotic electrical circuits, and are known
as multiple basins of consistency~\cite{Yip-Uchida}.
For limit-cycle oscillators, theoretical analysis has yielded
quite a few quantitative results explaining synchronization,
  desynchronization, and multiple synchronized groups or clusters
exhibited in an ensemble of limit-cycle oscillators~\cite{Teramae-Dan,
  Nagai-Nakao, Goldobin-Pikovsky, Nakao-Arai, Arai-Nakao, Nakao-Arai-Kawamura, Galan2}.

Our previous work~\cite{Arai-Nakao,Nakao-Arai} analyzed the linear
stability of synchronized or clustered states of uncoupled limit-cycle
oscillators subject to random common external impulses by calculating
the Lyapunov exponent, which quantifies the average rate of growth of
an infinitesimal phase separation between a pair of oscillators.
The only dynamical information we require about the oscillator is
contained in a simple function called the phase response curve (PRC)
describing the magnitude of phase advance or retardation due to a
perturbation at a given phase~\cite{Winfree,Kuramoto}.
The PRC has been measured in many oscillator-like systems, including
neurons, circadian oscillators, cardiac cells, and electrical
circuits~\cite{Tateno-Robinson, Galan-Ermentrout-Urban, Kobayashi,
  Gray, Arai-Nakao}.
For non-frequent impulses, the Lyapunov exponent $\Lambda$ is
given by
\begin{align}
  \Lambda = \lambda \int_{0}^{1} d\phi \int_{\bm{c}} d\bm{c} \ln
  \left| 1 + \frac{\partial}{\partial \phi} G(\phi, \bm{c}) \right|
  p(\bm{c}),
\end{align}
where $\lambda$ is the mean number of impulses in a unit time (or
rate), $G(\phi, \bm{c})$ is the PRC for an impulsive perturbation
whose intensity and direction (or {\em mark}~\cite{Hanson}) is
$\bm{c}$, $p(\bm{c})$ is the probability density of the mark, and the
integral is over the oscillator phase $\phi$ and the mark $\bm{c}$.  A
negative (positive) $\Lambda$ means that an infinitesimal phase
difference shrinks (grows) on the average, resulting in
synchronization (desynchronization) of the oscillators.

However, the Lyapunov exponent alone is not sufficient in
characterizing the whole coherence phenomena induced by the common
impulses, because it is an average quantity over the entire limit
cycle that characterizes only the local linear stability of the
synchronized state.
The phase difference generally does not monotonically decrease or
increase over successive common impulses due to fluctuations in the
expansion rates of the phase difference, which is determined by the
precise form of the PRCs.
When small external noises or inhomogeneities exist, such fluctuations
may induce large excursions from the synchronized state even if the
Lyapunov exponent is negative on average.
Oscillator pairs may find themselves with large phase difference, but
the global distribution of the phase difference cannot be explained by
a linear stability analysis.

In this paper, we further the theoretical analysis for an ensemble of
generic uncoupled limit-cycle oscillators to obtain the stationary
distribution of pair-wise phase difference \footnote{For an
  ensemble of uncoupled oscillators, no many-body effects due to
  coupling arise, and analyzing the phase relation between 2
  oscillators is sufficient to understand the situation for $N$
  oscillators.}.
Starting from general dynamical equations for a pair of limit-cycle
oscillators driven by common impulses, we derive a pair of random maps
and the corresponding two-body Frobenius-Perron
equation~\cite{Ott,Mackey} using the phase reduction
method~\cite{Winfree,Kuramoto,Arai-Nakao}.
We then derive an approximate one-body Frobenius-Perron
equation for the phase difference by averaging out the fast phase
dynamics, which yields the stationary distribution of the phase
difference.  The theoretical result is compared with direct numerical
simulations using FitzHugh-Nagumo oscillators receiving common Poisson
impulses.

\section{Theory}

\subsection{Phase reduction of the dynamical equation}

We investigate a pair of uncoupled oscillators receiving common random
impulses and also subject to a weak additive Gaussian white noise
independently.
The stochastic dynamical equation for the $i$-th oscillator in this pair is~\cite{Arai-Nakao}
\begin{equation}
  \label{OriginalODE}
  \dotXat{t} = \bm{F}(\Xa)
  + \sum_{n = 1}^{N(t)} \bm{\sigma}(\Xa, \bm{c}^{(n)}) h(t - t^{(n)})
  + \sqrt{D}\bm{H}(\Xa)\bm{\eta}_i,
\end{equation}
where $i = 1, 2$, $\Xat{t} \in \bm{R}^M$ is the oscillator state at
time $t$, $\bm{F} (\Xa):\bm{R}^M \to \bm{R}^M$ the dynamics of a
single oscillator, $N(t)$ the number of received impulses up to time
$t$, ${t^{(n)}}$ the arrival time of the $n$-th impulse, $\bm{c}^{(n)}
\in \bm{R}^K$ the intensity and direction, or {\em
  mark}~\cite{Hanson}, of the $n$-th impulse, $\bm{\sigma}(\Xa,
\bm{c}):\bm{R}^M \times \bm{R}^K \to \bm{R}^M$ is the coupling
function describing the effect of an impulse ${\bm c}$ to ${\bm
  X}_{i}$, $h(t - t^{(n)})$ is the infinitesimally narrow unit impulse
whose waveform is localized at the
time $t^{(n)}$ of the impulse
($\int^{\infty}_{-\infty} h(t - t^{(n)}) dt = 1$),
$\bm{H}(\Xa) \in \bm{R}^{M \times M}$ the coupling
matrix of the independent noise to the oscillator,
$\bm{\eta}_i \in \bm{R}^M$ a Gaussian white noise of unit intensity
with correlation
$\langle \eta_i^{\alpha}(t) \eta_j^{\beta}(s) \rangle = \delta(t -
s)\delta_{\alpha\beta}\delta_{ij}$ added independently to each
oscillator, and $D$ the intensity of the independent noise.
We interpret Eq.~(\ref{OriginalODE}) in the Stratonovich sense.
If the impulses and the independent noises are absent
($\bm{H} = \bm{0}$, $\bm{\sigma} = \bm{0}$),
the system is assumed to have a single stable
limit-cycle solution, ${\bm{X}_0(t)}$. 

As in our previous papers~\cite{Nakao-Arai,Arai-Nakao}, we use the
phase reduction method to analyze the dynamics of impulse-driven
oscillators.
We define an asymptotic phase~\cite{Winfree,Kuramoto} $\phi$ along the
limit cycle $\bm{X}_0(t)$ that constantly increases with a natural
frequency $\omega$, and extend the definition of phase to the whole
state space of the oscillator (except phase singular sets) by
identifying the orbits that asymptotically converge to the same point
on the limit cycle.
This defines a mapping from the oscillator state ${\bm X} \in {\bm
  R}^{M}$ to the phase $\phi \in [0, 1]$.

We assume that the interval between impulses is long compared
to the relaxation time back to the limit-cycle, so the oscillator is almost
always on the limit-cycle when an impulse is received.
We can then reduce Eq.~(\ref{OriginalODE}) to the dynamics
of a single asymptotic phase $\phi_i$.
The dynamics of the phase $\phi_i^{(n)}$ right before the $n$-th
impulse is received can be approximately described by a random
map
\begin{equation}\label{RandomMap}
  \phi_i^{(n+1)} = \phi_i^{(n)} + G(\phi_i^{(n)}, \bm{c}^{(n)})
  + \omega \tau^{(n)} + \gamma_{i}^{(n)},
\end{equation}
where $G(\phi, \bm{c})$ is the PRC, $\omega \tau^{(n)}$
  is the increase in phase
  during
  the interval between the $n$-th and $(n+1)$-th impulses $\tau^{(n)}
= t^{(n+1)} - t^{(n)}$, and $\gamma_{i}^{(n)}$ is the
displacement caused by the additive independent Gaussian noise
$\bm{\eta}_{i}$ in the interval $\tau^{(n)}$.  
From now on, we assume that the range of $\phi$ to be the real numbers
${\bm R}$ by taking into account the number of windings around the
limit cycle, which makes the treatment of periodic boundary conditions
easier in the following derivation~\cite{Ermentrout-Saunders}.  

The PRC $G(\phi, \bm{c})$ describes the change in phase of the
oscillator when an impulse of mark $\bm{c}$ is received at phase
$\phi$ on the limit cycle, which is periodic in $\phi$, i.e.
$G(\phi+1, \bm{c}) = G(\phi, \bm{c})$.
It can be obtained by applying the approximation theorem by
Marcus~\cite{Marcus} to the impulsive term in Eq.~(\ref{OriginalODE})
as~\cite{Arai-Nakao}
\begin{align}
  G(\phi, \bm{c}) = \phi\left( {\bm X}_{0}(\phi) + {\bm g}({\bm
      X}_{0}(\phi), \bm{c}) \right) - \phi,
\end{align}
where ${\bm g}({\bm X}, \bm{c}) =\left\{ \exp\left( \sum_{j}
    \sigma_{j}({\bm X}, \bm{c}) ( \partial / \partial X_{j} ) \right)
  - 1 \right\} {\bm X}$
\footnote{For the Ito interpretation of the impulse term,
  the PRC is simply given by $ G(\phi, \bm{c}) = \phi({\bm
    X}_{0}(\phi) + {\bm \sigma}({\bm X}_{0}(\phi), \bm{c})) -
  \phi$~\cite{Arai-Nakao}.}.
The PRC is related to the phase sensitivity function~\cite{Kuramoto}
$Z_i(\phi) \equiv \left. \partial \phi / \partial X_i \right|_{{\bm X}
  = {\bm X_{0}}(\phi)}$
by $G(\phi, \bm{c}) \simeq {\bm Z}(\phi) \cdot \sigma({\bm
  X}_{0}(\phi), \bm{c})$ when the effect of the impulse $\sigma({\bm
  X}_{0}(\phi), \bm{c})$ is small.

Generally speaking, the displacement $\gamma_{i}^{(n)}$ depends on the
oscillator phase $\phi_{i}^{(n)}$, the impulse mark ${\bm c}^{(n)}$,
and the relaxation path to the limit cycle after each impulse.
We approximate the actual distribution function of $\gamma_{i}^{(n)}$
by a zero-mean Gaussian normal distribution with variance $\epsilon^2
\tau^{(n)}$   
\footnote{The Stratonovich interpretation of Eq.~(\ref{OriginalODE}) introduces a phase-dependent drift term that disappears
  upon averaging over the limit-cycle~\cite{Nakao-Arai-Kawamura}, so the additive diffusion term $\gamma_i^{(n)}$ may be taken to be zero mean.}.  The approximate diffusion constant $\epsilon$ can be obtained by
ignoring the fast relaxation dynamics to the limit cycle after the
impulse and by averaging the phase dependence over the limit cycle
as~\cite{Nakao-Arai-Kawamura}
\begin{equation}
  \epsilon^2
  =
  \int_{0}^1 \sum_{ijk} Z_i(\phi) Z_j(\phi)
  H(\bm{X}_0(\phi))_{ik} H(\bm{X}_0(\phi))_{jk} d\phi,
\end{equation}
where we utilize the fact that the stationary phase distribution of a
single oscillator receiving infrequent impulsive forcing is nearly
uniform~\cite{Arai-Nakao,Nakao-Arai}.
As we demonstrate later, this is a
  good approximation for oscillators whose relaxation to the
  limit cycle is sufficiently fast.

\subsection{Frobenius-Perron equation for the phase difference}

Let us consider the dynamics of the joint probability
  distribution $\rho(\phi_1, \phi_2, n)$ of the phases $(\phi_1,
  \phi_2)$ right before the $n$-th impulse, determined by the random
map Eq.~(\ref{RandomMap}).
We assume the range of phase variables to be $\phi_{1,2} \in \bm{R}$.
The Frobenius-Perron equation for the evolution of the joint
  distribution is
\begin{eqnarray} 
  \label{FPE0}
&&
\rho(\phi_1, \phi_2, n + 1)
\cr \cr
&=&  
\int_{-\infty}^{\infty} d\phi_1'
\int_{-\infty}^{\infty} d\phi_2' 
\int_{0}^{\infty} d\tau
\int_{\bm{c}} d{\bm c}
\int_{-\infty}^{\infty} d\gamma_{1}
\int_{-\infty}^{\infty}d\gamma_{2}
W(\tau)
p({\bm c})
R(\gamma_{1}, \tau) R(\gamma_{2}, \tau)
\ \times 
\cr \cr
&&
\delta\bigl(\phi_1 - \phi_1' - G(\phi_1',{\bm c}) - \omega\tau - \gamma_{1}\bigr)
\delta\bigl(\phi_2 - \phi_2' - G(\phi_2',{\bm c}) - \omega\tau - \gamma_{2}\bigr)
\rho(\phi_1', \phi_2', n)
\cr \cr
& = &
\int_{-\infty}^{\infty} d\phi_1'
\int_{-\infty}^{\infty} d\phi_2'
\int_{0}^{\infty} d\tau
\int_{\bm{c}} d{\bm c}
W(\tau)
p({\bm c})
R\bigl(\phi_1 - \phi_1' - G(\phi_1',{\bm c}) - \omega\tau, \tau\bigr)
\ \times
\cr \cr
&&
R\bigl(\phi_2 - \phi_2' - G(\phi_2',{\bm c}) - \omega\tau, \tau\bigr)
\rho(\phi_1', \phi_2', n),
\end{eqnarray}
where $W(\tau)$ is the inter-impulse distribution, $G(\phi,{\bm
  c})$ is the PRC, and $R(\gamma_{i}, \tau)$ is the probability
that an oscillator $i$ has diffused an amount $\gamma_{i}$ in a time
interval $\tau$,
which we approximated as a normal distribution with variance
$\epsilon^2 \tau$.

Going to the center-of-mass coordinates, we change variables to $\psi
= ( \phi_1 + \phi_2 ) / 2$ and $\xi = \phi_1- \phi_2$, where
  $\psi$ is the mean phase and $\xi$ is the phase difference.
The Frobenius-Perron equation~(\ref{FPE0}) is transformed as
\begin{eqnarray}\label{ApproxDiffusion} \nonumber
  &&
  \rho(\psi, \xi, n + 1) 
  =
  \int_{-\infty}^{\infty} d\psi'
  \int_{-\infty}^{\infty} d\xi' 
  \int_{0}^{\infty} d\tau
  \int_{\bm{c}} d{\bm c}
  p({\bm c})
  W(\tau)
  \times
  \cr \cr
  &&
  R\left(\psi + \frac{\xi}{2} - \psi' - \frac{\xi'}{2} - G\left(\psi' + \frac{\xi'}{2}, {\bm c}\right) - \omega \tau, \tau \right)
  \times \cr \cr
  &&
  R\left(\psi - \frac{\xi}{2} - \psi' + \frac{\xi'}{2} - G\left(\psi' - \frac{\xi'}{2}, {\bm c}\right) - \omega \tau, \tau \right)
  \rho(\psi', \xi', n).
\end{eqnarray}

We now restrict the mean phase to $\psi \in [0, 1)$ and the phase
difference to $\xi \in (-1, 1)$ similarly to Ermentrout and
Saunders~\cite{Ermentrout-Saunders} by introducing a new distribution
function
\begin{equation}
  P(\psi, \xi, n) = \sum_{p=-\infty}^{\infty}
  \sum_{q=-\infty}^{\infty} \rho(\psi + p, \xi + 2q, n),
\end{equation}
which sums up contributions from pairs of phase values with different
winding numbers but represent physically equivalent situations on the
limit cycle.
This "wrapped" $P(\psi, \xi, n)$ corresponds to the actual
  distribution of the mean phase and the phase difference measured in
  simulations or experiments.
Using the periodicity of the PRC, we obtain
\begin{eqnarray}\label{WrappedFP} \nonumber
&&
P(\psi, \xi, n + 1)
=
\sum_{\pi(p) = \pi(q)}
\int_{0}^1 d\psi'
\int_{-1}^1 d\xi' 
\int_{0}^{\infty} d\tau
\int_{\bm{c}} d{\bm c}
p({\bm c})
W(\tau)
\times
\cr\cr
&&
R\left(\psi +\frac{\xi}{2} - \psi'  - \frac{\xi'}{2} + p - G\left(\psi' + \frac{\xi' }{2}, {\bm c} \right) - \omega \tau, \tau \right)
\times 
\cr \cr 
&&
R\left(\psi - \frac{\xi}{2} - \psi' + \frac{\xi'}{2} + q - G\left(\psi' - \frac{\xi'}{2}, {\bm c} \right) - \omega \tau, \tau \right) P(\psi', \xi', n),
\end{eqnarray}
where the summation involves all pairs of $p$ and $q$ of equal parity
($\pi(\cdot)$ denotes the parity of an integer).

To obtain a closed equation for the phase difference $\xi$, we now
average out the fast dynamics of the mean phase, $\psi$.
If the impulses are not so frequent and the magnitude of the
independent noise is small, the mean phase $\psi$ is a rapidly
changing variable compared to the phase difference $\xi$.
Then $\psi$ and $\xi$ can be taken to be nearly independent,
and the joint probability density can be separated as
$P(\psi, \xi, n) \simeq S(\psi, n) U(\xi, n)$,
where $S(\psi, n)$ and $U(\xi, n)$ are the probability density
  functions of $\psi$ and $\xi$, respectively.  Note that $U(\xi, n)$
  is periodic in $\xi$, $U(\xi \pm 1, n) = U(\xi, n)$, because $\xi$ and
  $\xi \pm 1$ represent the same phase difference. 
For non-frequent impulses, $\psi$ is almost uniformly
distributed on the limit cycle, $S(\psi, n) \simeq
1$~\cite{Nakao-Arai,Arai-Nakao}.
We then average over the $\psi$ on both sides to obtain
\begin{equation}
  \label{psixiEq}
  U(\xi, n + 1) =
  \int_{-1}^1 d\xi'
  \int_0^1 d\psi'
  \int_0^1 d\psi
  T(\psi, \xi, \psi', \xi')
  U(\xi', n),
\end{equation}
where
\begin{eqnarray}\label{TEq}
&&
T(\psi, \xi, \psi', \xi')
=
\sum_{\pi(p) = \pi(q)}
\int_{0}^{\infty} d\tau
\int_{\bm{c}} d{\bm c}
p({\bm c})
W(\tau)
\times
\cr \cr
&&
R\left(\psi +\frac{\xi}{2} - \psi'  - \frac{\xi'}{2} + p - G\left(\psi' + \frac{\xi' }{2}, {\bm c} \right) - \omega \tau, \tau \right)
\times 
\cr \cr
&&
R\left(\psi - \frac{\xi}{2} - \psi' + \frac{\xi'}{2} + q - G\left(\psi' - \frac{\xi'}{2}, {\bm c} \right) - \omega \tau, \tau \right).
\end{eqnarray}
We now derive an approximate one-body Frobenius-Perron
equation for the distribution of the phase difference
\begin{equation}\label{Eq:Iterative}
  U(\xi, n + 1) = \int_{-1}^1 
  X(\xi, \xi') U(\xi', n) 
  d\xi',
\end{equation}
where the transition probability is given by
\begin{equation}
  X(\xi, \xi') =
  \int_0^1 d\psi'
  \int_0^1 d\psi
  T(\psi, \xi, \psi', \xi').
\end{equation}
Namely, we have reduced the problem to finding the stationary
distribution of a Markov process for the random variable $\xi$ with
transition probability $X(\xi, \xi')$.
By numerically estimating the transition probability $X(\xi,
  \xi')$ from the PRC, Eq.~(\ref{Eq:Iterative}) can be iterated until
a stationary state is reached.
$X(\xi, \xi')$ is periodic in $\xi$ and $\xi'$, $X(\xi \pm 1,
  \xi' \pm 1) = X(\xi, \xi')$.

In the following numerical simulations, we assume that the
  random impulses are generated by a Poisson process, and fix $\bm{c}$
  so that all impulse marks are identical.  The inter-impulse interval
  is exponentially distributed,
\begin{equation}
  W(\tau) = \frac{1}{\tau_{P}} \exp \left( - \frac{\tau}{\tau_P} \right),
\end{equation}
where the parameter $\tau_{P}$ is the mean impulse interval.
We further simplify the calculation by neglecting the dependence of
$R(\gamma_{i}, \tau)$ on $\tau$ in Eq.~(\ref{TEq}) by replacing it
with $R(\gamma_{i}, \tau_p)$, a normal distribution with fixed
variance $\epsilon^2 \tau_P$, which is equal to the average variance
of the diffusion $\gamma_{i}$ in a mean inter-impulse interval
$\tau_P$.
Defining $G_-' = G(\psi' + \xi'/2, {\bm c}) - G(\psi' - \xi' / 2, {\bm c})$ and $G_+' =
G(\psi' + \xi'/2, {\bm c}) + G(\psi' - \xi' / 2, {\bm c})$, the function $T(\psi, \xi,
\psi', \xi')$ can then explicitly be calculated as
\begin{eqnarray}
&&
T(\psi, \xi, \psi', \xi')
=
\frac{\exp\left(D / 4\tau_P \omega^2\right)}{\omega\tau_P}
\sqrt{\frac{D \tau_P}{4\pi}} 
\sum_{p\;\mbox{\small even}}
\exp\left(
	-\frac{(\xi - \xi' - G_-' + p)^2}{4 D \tau_P}
	\right) \times
\cr \cr
&&
\sum_{q}
\exp \left(
	-\frac{\psi - \psi' - G_+'/2 + q}{\omega\tau_P}
	\right)
\left(
	\mbox{erf}
	\left(\frac{2\omega(\psi - \psi' - G'_+ / 2 + q) - D}{2\omega\sqrt{D\tau_P}}\right) + 1\right),
\end{eqnarray}
where $\mbox{erf}$ is the Gauss error function.  In numerical
calculations, using the first several terms in the summation for $p$
is sufficient.  Since the error function approaches $1$ ($-1$) very quickly for positive (negative) values of its argument, for a small enough value of $D$, the sum over $q$ is to a good approximation a geometric series.

\section{Numerical Simulations}

As an example of a limit-cycle oscillator, we employ the
FitzHugh-Nagumo (FHN) neural oscillator~\cite{Koch} driven by common
Poisson impulses and independent Gaussian-white noises described by
the following set of equations:
\begin{eqnarray}
\label{FHN}
\uadt & = & \varepsilon (\va  + a - b \ua), \cr 
\vadt & = & \va - \frac{\vaq}{3} - \ua + I_0 + \sigma(\va, c) \sum_{n=1}^{N(t)} h(t - t_{n}) + \sqrt{D}\eta_i(t).
  \;\;\;\;\;\;\;\;\;
\end{eqnarray}
Here, parameters $\varepsilon, a, b$ are fixed at $\varepsilon =
0.08$, $a = 0.7$, $b = 0.8$, and we use the parameter $I_0$ as a
bifurcation parameter.
The last two terms of the equation for $v$ describe impulses and
noises, where $h(t)$ represents a unit impulse and $\sigma(v, c)$
describes $\va$-dependent effect of the impulse to the oscillator.
In this example, both $\bm{H}$ and $\bm{\sigma}$ have only one non-zero
component.
For simplicity, we take the impulse strength $c$ to be a constant value.  
When both terms are zero, a limit cycle exists for $I_0 \in
[0.331, 1.419]$, which is created by a subcritical Hopf bifurcation at
either limits of $I_0$.  For the simulations, we employ $I_0 = 0.34$
and $I_0 = 0.875$, which give oscillator periods of $T \simeq 46.792$
and $T \simeq 36.418$, respectively.
We choose these values because the oscillator characteristics change
in such a way as to show synchronized and desynchronized states for
additive impulses, and stable 2-cluster states for linear
multiplicative impulses.
We set the mean interval between the impulses at $\tau_P = 10T$.
Results similar to the following have been obtained using
Stuart-Landau and Moris-Lecar oscillators. However, we restrict our
discussion to the FitzHugh-Nagumo model as it displays all of the
salient features of interest.

In direct numerical simulations of Eq.~(\ref{FHN}), we realized the Stratonovich
interpretation by using a colored Gaussian
noise generated by the Ornstein-Uhlenbeck process $\tau
\dot{\eta}(t) = -\eta(t) + \chi(t)$, where $\chi(t)$ is a Gaussian white
noise of unit intensity, and delivering the impulses as discontinuous
jumps of amplitude given by the Marcus approximation theorem of
continuous physical jumps~\cite{Arai-Nakao,Marcus}.  The correlation
time $\tau$ of $\eta(t)$ was set to $0.05$, which is much shorter than
the oscillator period $T$.
In calculating the Frobenius-Perron equation~(\ref{Eq:Iterative}), we numerically estimate
$X(\xi, \xi')$ and $U(\xi)$ on discrete grids of dimensions between
$128$ to $2048$ for $\xi$ and $\xi'$, depending on how rapidly $X(\xi,
\xi')$ varies as a function of $\xi$ and $\xi'$.  Generally, the
larger the value of $D$, the lower the required resolution.

\begin{figure}[htbp]
  \centering
 \includegraphics*[width=0.95\hsize]{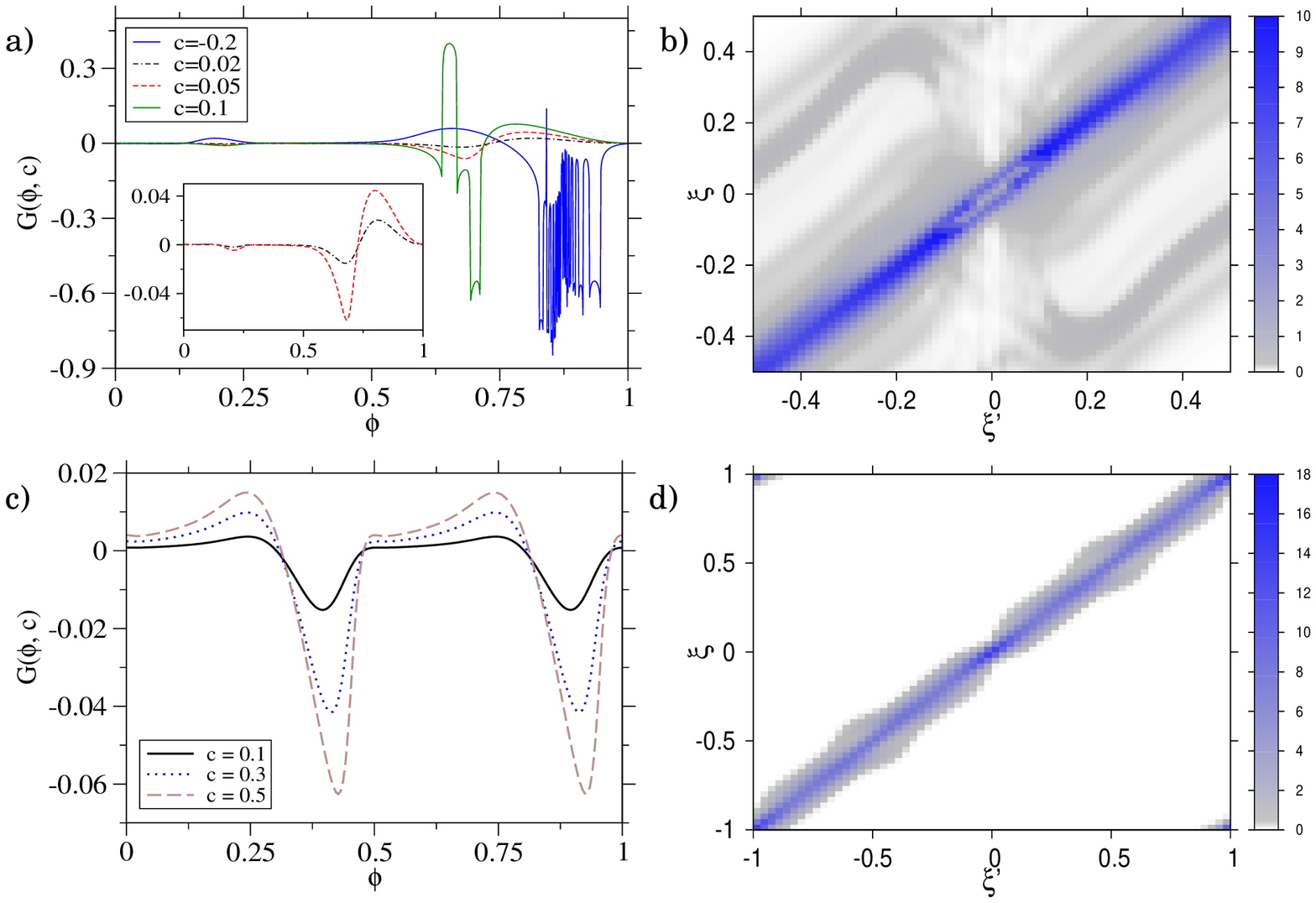}
  \caption{(Color online) a) The PRC $G(\phi)$ for various values of
    additive impulse intensity $c$ for the FHN oscillator with $I =
    0.34$, with the PRCs of smaller amplitudes shown enlarged in the
    inset.  b) The averaged phase difference transition probability
    $X(\xi, \xi')$ for additive impulses with $c = -0.2, D = 2.5
    \times 10^{-5}$, corresponding to the case shown in
    Fig.~\ref{Fig:Synch}.  c), d) The PRCs for $I = 0.875$ with
    multiplicative impulses, and the corresponding transition
    probability for $c = 0.5, D = 2.5 \times 10^{-5}$,
    corresponding to the case shown in Fig.~\ref{Fig:Clusters}.  The
    PRC of FHN gains additional symmetry $G(\phi) = G(\phi + 0.5)$
    (as does the transition probability $X(\xi, \xi') = X(\xi \pm 0.5, \xi' \pm 0.5)$) with application of balanced, multiplicative noise, $\sigma(v, c)
    = cv$.}
  \label{Fig:GandX}
\end{figure}

We show examples of PRCs for different values of the impulse strength
$c$ obtained for the FHN oscillator through simulation in
Fig.~\ref{Fig:GandX}, as well as the resultant transition probability
$X(\xi, \xi')$.
In all of the figures, we only show $\xi \in [-0.5, 0.5]$ as
  $X(\xi, \xi')$ and $U(\xi)$ are periodic.
The Lyapunov exponent $\Lambda$ is negative for the smooth PRCs, and
positive for the rapidly fluctuating PRCs.
The generic dynamical behavior of the oscillators are as
follows~\cite{Arai-Nakao}: When $\Lambda < 0$, the system settles down
into a largely quiescent state once synchronization is
  achieved.
The rare but sudden disintegration of a pair of oscillators is
possible if there are regions of the PRC with positive local Lyapunov
exponent, but the relative separation of a pair remains largely static.
However for $\Lambda > 0$, disintegration of a pair happens
routinely, followed by a gradual reunion, and this cycle continues
{\it ad infinitum}.  These occasional sudden, large excursions from the
synchronized state is generally known as modulational or on-off
intermittency~\cite{Fujisaka-Yamada,Pikovsky}, and is a characteristic
behavior of a random multiplicative process, of which our system is an
example.

\begin{figure}[htbp]
  \centering
 \includegraphics*[width=0.95\hsize]{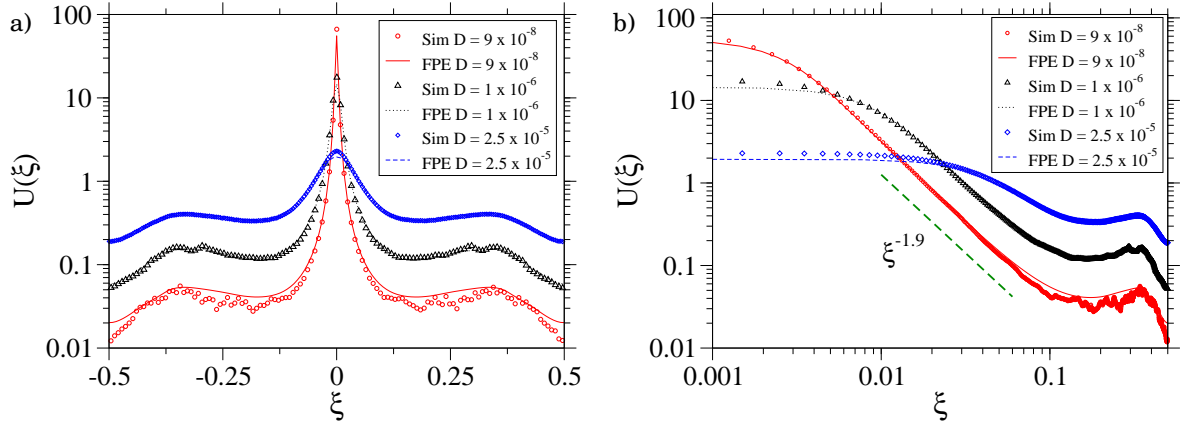}
  \caption{(Color online) Comparison of $U(\xi)$ for the case of
    $\Lambda < 0$ calculated using the averaged Frobenius-Perron
    equation (FPE) and measured via simulation (Sim).  a) shows the
    global distribution in semi-log scales, and b) shows the
    distribution near $\xi = 0$ in log-log scales for $\xi > 0$.  The
    intensity of independent, additive noise (diffusion) is varied ($D = 9
    \times 10^{-8}$, $1 \times 10^{-6}$, $2.5 \times 10^{-5}$) while the
    intensity of the common impulses ($c = 0.5$) is kept constant for
    FHN oscillators with $I_0 = 0.875$.  It can be seen that lowering
    the independent noise narrows and increases the height of the
    peaks of the distribution near $\xi = 0$.  Because the Lyapunov
    exponent remains constant, the slope is preserved for various
    diffusion strengths.}
  \label{Fig:Synch}
\end{figure}

\begin{figure}[htbp]
  \centering
 \includegraphics*[width=0.95\hsize]{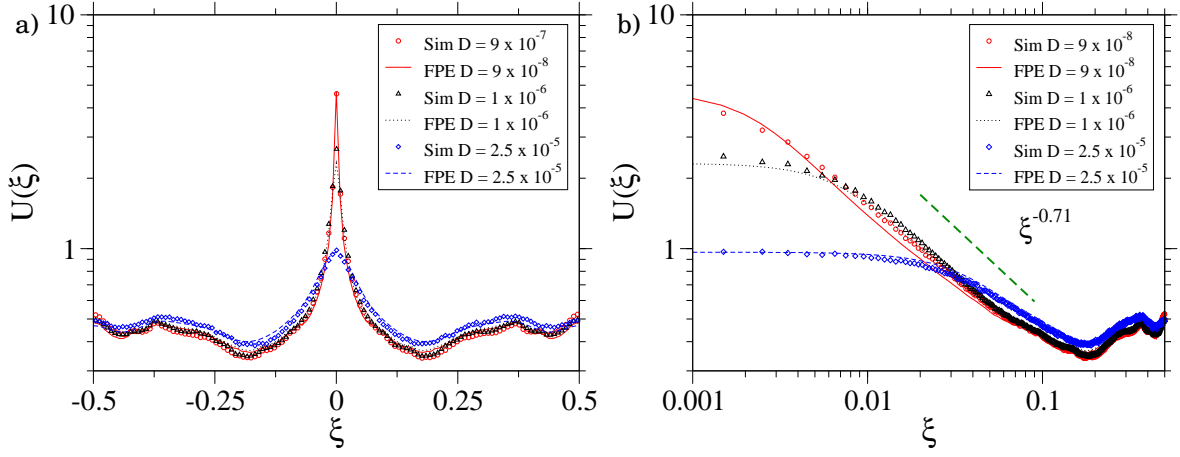}
  \caption{(Color online) Comparison of $U(\xi)$ for the case of
    $\Lambda > 0$ calculated using the averaged Frobenius-Perron
    equation (FPE) and measured via simulation (Sim).  a) shows the
    global distribution in semi-log scales (note the y-axis range in
    comparison with Fig.~\ref{Fig:Synch} and Fig.~\ref{Fig:Clusters}),
    and b) shows the distribution near $\xi = 0$ in log-log scales.
    The intensity of independent, additive noise is varied ($D =
    9 \times 10^{-8}$, $1 \times 10^{-6}$, $2.5 \times 10^{-5}$) while
    the intensity of the common impulses ($c = -0.2$) is kept constant
    for FHN oscillators with $I_0 = 0.34$. Due to the inherent
    instability of the $\xi = 0$ state, the distribution of $\xi$
    reaches a limiting value as the independent, additive noise is
    lowered.}
  \label{Fig:Desynch}
\end{figure}

\begin{figure}[htbp]
  \centering
 \includegraphics*[width=0.8\hsize]{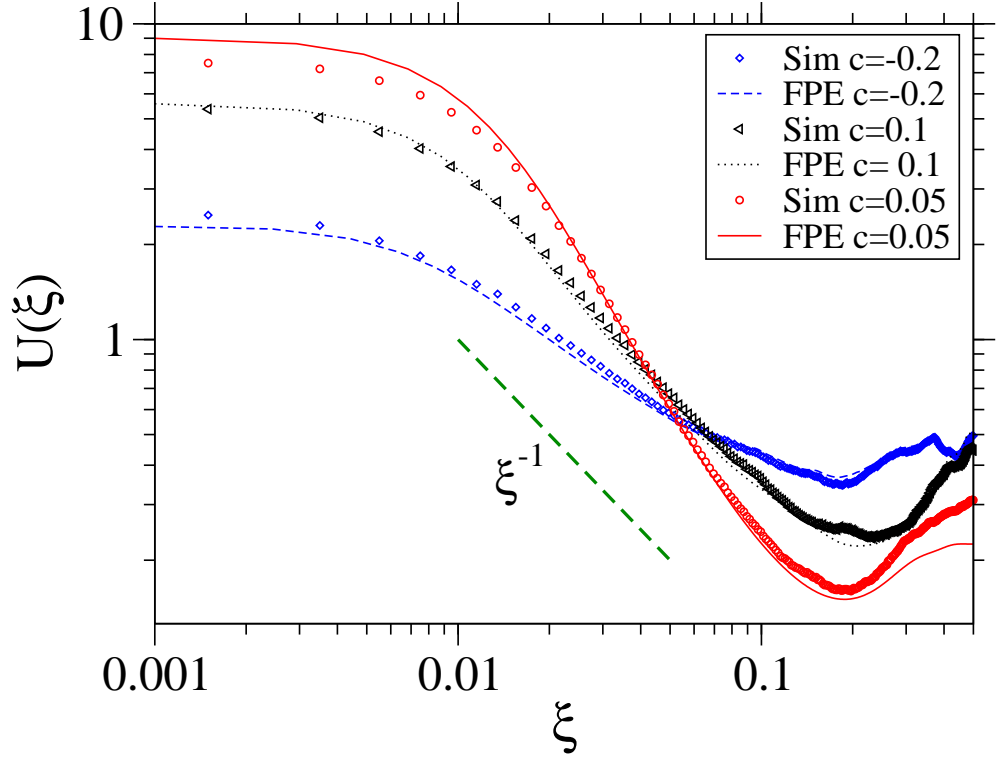}
  \caption{(Color online) Power-law distributions of phase difference
    $U(\xi)$ near $\xi=0$ in log-log scales for the FHN oscillator with $I = 0.34$.  The
    intensity of independent, additive noise is kept constant ($D
    = 1 \times 10^{-6}$) while the intensity of the common impulses
    are varied ($c = -0.2, 0.05, 0.1$).  As the Lyapunov exponent of
    the system is changed, the slope of the power-law changes
    correspondingly.}
  \label{Fig:Lyapunov}
\end{figure}

Now let us examine the stationary distribution $U(\xi)$ of the
  phase difference $\xi$.
We expect the distribution of $\xi$ to be qualitatively different
between $\Lambda$ of different sign.
Figures~\ref{Fig:Synch} and ~\ref{Fig:Desynch} show the distribution
of $\xi$ for additive impulses ($\sigma(v, c) \equiv c$, $c = 0.5,
-0.2$, respectively) at various intensities of independent noise for
PRCs with negative and positive $\Lambda$.
In all figures, theoretical curves obtained using our
  Frobenius-Perron equation for the phase difference nicely fit the
  results of direct numerical simulations, which indicates that the
  approximations we have made so far are reasonable for the parameter
  values we use.
It is readily apparent that if the synchronized state is stable, the
synchronized peaks become taller and narrower as the diffusion is made
smaller, while $\xi$ far away from the stable peaks become
increasingly rare.
On the other hand, if the synchronized state is unstable, the
distribution for rare $\xi$ reaches a limiting value, while only the
tip of the synchronized peak increases in height and the width of the
peak remains constant.
The distributions exhibit a power-law dependence near $\xi = 0$, a
characteristic of random multiplicative
processes~\cite{Fujisaka-Yamada,Pikovsky,NakaoRMP,Kitada}.
As shown in Fig.~\ref{Fig:Lyapunov}, different power-law
  exponents are obtained by changing the impulse strength, $c$ ($=
  -0.2, 0.05, 0.1$), where the Lyapunov exponent $\Lambda$ determines
  whether the slope of the power law is steeper or shallower than
  $-1$~\cite{Fujisaka-Yamada,Pikovsky,NakaoRMP,Kitada}.

\begin{figure}[htbp]
  \centering
  \includegraphics*[width=0.95\hsize]{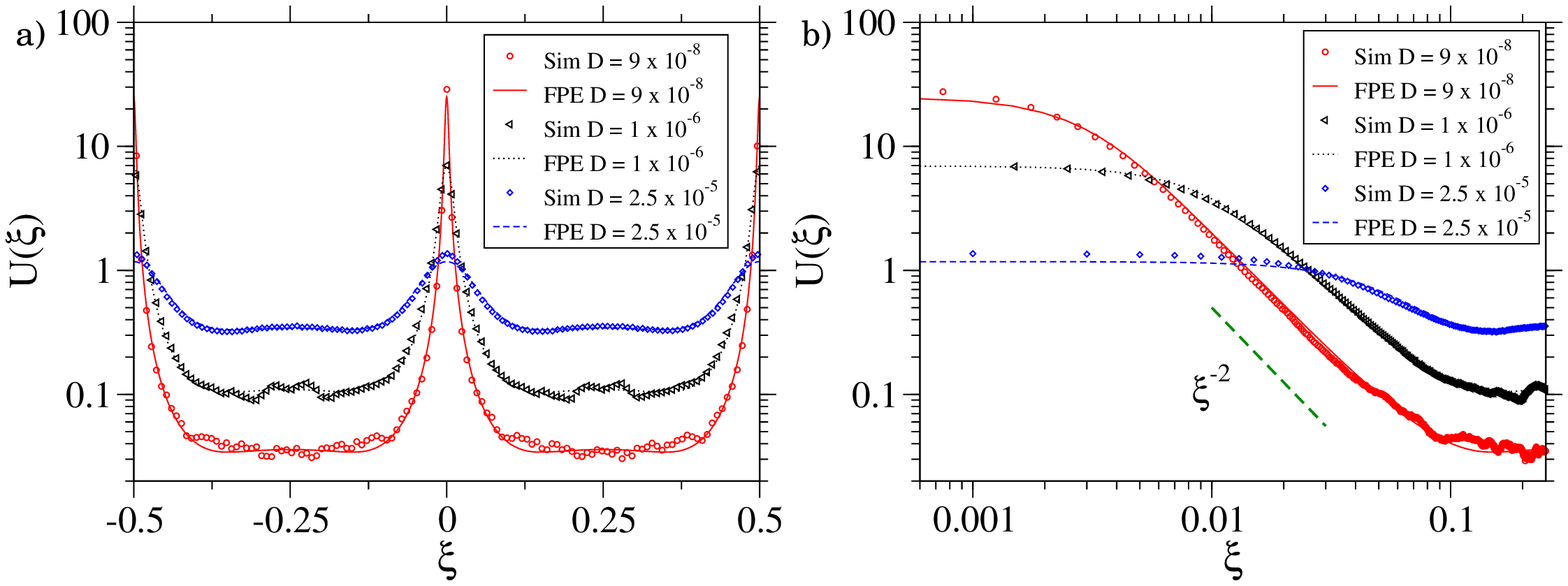}
  \caption{(Color online) Comparison of 2-clustered $\xi$ distribution
    for the case of $\Lambda < 0$ calculated using the averaged
    Frobenius-Perron equation (FPE) and measured via simulation (Sim)
    for impulses with $c = 0.5$, FHN bifurcation parameter $I_0 =
    0.875$ and independent additive noise ($D = 9 \times
    10^{-8}$, $1 \times 10^{-6}$, $2.5 \times 10^{-5}$). a) shows the
    global distribution in semi-log scales, and b) shows the distribution near $\xi = 0$ in log-log scales.}
  \label{Fig:Clusters}
\end{figure}

Figure~\ref{Fig:Clusters} shows the same basic mechanism at work for
the case with linear multiplicative impulses ($\sigma(v, c) = cv$, $c
= 0.5$), which exhibits symmetric 2-cluster states.
The distribution, which is nicely fitted by the theoretical curve, has
three peaks in this case, corresponding to the three
  possible phase differences in the 2-cluster states ($\xi=0$ and $\xi
  = \pm 0.5$, where $\xi = +0.5$ and $\xi = -0.5$ represent the same
  phase difference).
Near each peak, the distribution exhibits power-law dependence as for
the case of additive impulses. 

\section{Comparison with coupled oscillators}

We have shown that common random impulses applied to a pair of
uncoupled limit-cycle oscillators generally produce phase coherence.
Much existing work focuses on the self-organizing coherence brought
about through coupled elements, so we would like to touch upon the
similarities and differences between the coherence observable between
coupled systems and uncoupled systems receiving a common random input.
For simplicity, we consider a pair of identical oscillators.

Sufficiently weak common random input to uncoupled oscillators always
tend to stabilize the synchronized state at zero phase difference
regardless of the shape of the PRC.  The probability density function
$U(\xi)$ of the phase difference $\xi$ always has a peak at $\xi = 0$,
as we have seen in Figs.~\ref{Fig:Synch}, \ref{Fig:Desynch} and \ref{Fig:Clusters}.  When the common input is stronger,
the in-phase synchronized state $\xi = 0$ can be unstable.  We
nevertheless observe that $U(\xi)$ has a local maximum at $\xi = 0$ as
shown in Fig. 3 for weakly unstable situations.  For much stronger
inputs, the PRC can take highly irregular forms that contain many
discontinuities or with many rapid, large amplitude oscillations.  It
is then possible for $U(\xi)$ to have a local minimum at $\xi = 0$.

In contrast, for oscillators with weak mutual coupling, the in-phase
synchronized state may either be stable or unstable depending on the
shape of the PRC and the interaction function between the oscillators.
If the in-phase state is unstable, there would be no peak appearing at
$\xi = 0$; instead, a peak would be expected at some other $\xi \neq
0$~\cite{Ermentrout-Saunders, Pfeuty}.

This illustrates the biggest difference between coherence in mutually
coupled systems and uncoupled systems subject to common inputs.  In
coupled systems, it is possible to have a single stable phase-locked
state with $\xi \neq 0$, while in uncoupled systems, this is not
possible.
One possible point of confusion that arises here may be our use of the
terms ``stable'' and ``unstable''.
For uncoupled oscillators driven by common input, these terms
represent {\it statistical} stability of the synchronized state.  Even
if the synchronized state induced by common input is slightly
unstable, distribution of the phase differences can still have a
shallow maximum at zero phase difference.  The vicinity of $\xi = 0$
is an attractive region even if the synchronized state is weakly
unstable.
In contrast, these terms represent {\it deterministic} stability for
coupled systems.  If it is unstable, we never observe such a maximum
even if independent noises are added.

If the natural frequencies of the oscillators are different, the
difference in phase coherence behavior will be more subtle.
In this case, a local extremum in $U(\xi)$ at $\xi \neq 0$ appears for two non-identical oscillators driven by common input, and
may be a maximum or minimum depending on the degree of statistical stability
or instability of the locked state (data not shown).  In weak mutually coupled systems, the deterministic
stability is once again dependent on the interaction function, and in addition,
the magnitude of the difference of the natural frequencies.  Furthermore, combined effects of coupling and common input, which may
be important in practical situations, will lead to more intriguing
behavior.

\section{Summary}

We have found an approximate method to calculate the steady-state
probability distribution of the pair-wise phase difference in an
ensemble of uncoupled oscillators receiving random impulses.  The
system is essentially a random multiplicative process, and as such
shows modulational intermittent behavior and power-law dependence of
the distribution near its peak.  Qualitative and quantitative features
of the distributions have been found relating the results to the
Lyapunov exponents that characterized the stability of clustered
states in earlier works~\cite{Arai-Nakao,Nakao-Arai}.

Our treatment is conceptually a generalization of our previous
result~\cite{Nakao-Arai-Kawamura} on uncoupled limit-cycle oscillators
subject to common and independent infinitesimal Gaussian-white noises.
In that case, the common noise always stabilizes the synchronized
state as long as an oscillator possesses a continuous phase
sensitivity function.  The oscillators form one or more synchronized
clusters, depending on the degree of symmetry possessed by the system.
By contrast, in the scenario studied in this paper, there is the
further possibility that common impulses may destabilize the
synchronized state, which can still quantitatively be analyzed within
our theoretical framework based on the averaged Frobenius-Perron
equation~\footnote{The slope of the power-law dependence of $U(\xi)$
  near the peak is always $-2$ for the infinitesimal Gaussian-white
  drive, while it can take a range of values in the present impulsive
  drive.}.

In this work, we considered a pair of identical oscillators subject to
the same common impulses, and considered the diffusion in between
received impulses as the effect of independent noises.  Our method can
also be applicable if the natural frequency or the PRC of the
oscillators are slightly different.  Furthermore, we can also
interpret the diffusion as the result of inherently noisy response of
an oscillator to pulsatile inputs.  The consequences of a noisy PRC
has been treated recently in the case of mutually coupled neural
oscillators~\cite{Ermentrout-Saunders}.  Mildly chaotic, non-mixing
oscillators also show a similar noisiness to their responses.  A noisy
PRC also arises in the case of globally coupled oscillators exhibiting
a collective coherent oscillation, where the response of the
collective oscillation is inherently fluctuating due to finite-size
effects, in particular near the critical point of synchronization
transition~\cite{KawamuraCollective}.  The method developed within
this paper may prove to be useful in analyzing the dynamics of such
systems.  Further results will be reported on in the near future.


This work is supported by the Grant-in-Aid for Young Scientists (B),
19760253, 2008, from MEXT, Japan.

\clearpage

\begin{thebibliography}{11}

\bibitem{Mainen-Sejnowski}
  Z. F. Mainen and T. J. Sejnowski,
  Science {\bf 268}, 1503 (1995).

\bibitem{Binder-Powers}
  M. D. Binder and R. K. Powers,
  J. Neurophysiol {\bf 86}, 2266 (2001).

\bibitem{Galan}
  R. F. Gal\'an, N. F. Trocme, G. B. Ermentrout, and N. N. Urban,
  J. Neurosci {\bf 26(14)}, 3646 (2006).

\bibitem{Roy} R. Roy and K.S. Thornburg, Jr., Phys. Rev. Lett. {\bf
    72}, 2009 (1994); A. Uchida, R. McAllister, and R. Roy, Phys.
  Rev.  Lett. {\bf 93}, 244102 (2004).
  
\bibitem{Yoshida} K. Yoshida, K. Sato, A.  Sugamaga, J. Sound and
  Vibration {\bf 290}, 34 (2006).

\bibitem{Yip-Uchida} H. Yip, S. Sano, A. Uchida and S. Yoshimori,
  ``Multiple basins of consistency in a Mackey-Glass electronic
  circuit driven by colored noise'' in Proceedings of NOLTA 2007,
  Vancouver, Canada (2007).
  
\bibitem{Arai-Nakao} 
  K. Arai and H. Nakao,
  Phys. Rev. E {\bf 77}, 036218 (2008).
  
\bibitem{Hudson} Y. Zhai, I. Z. Kiss, P. A. Tass, and J. L. Hudson,
  Phys. Rev. E {\bf 71}, 065202(R) (2005).
  
\bibitem{Kobayashi} H. Ukai, T. J. Kobayashi, M. Nagano, K. Masumoto,
  M. Sujino, T. Kondo, K. Yagita, Y. Shigeyoshi and H. R. Ueda, Nature
  Cell Biology {\bf 9}, 1327 (2007).
  
\bibitem{Toral} R. Toral, C. R. Mirasso, E. Hern\'andez-Garc\'ia, and
  O. Piro, Chaos {\bf 11}, 665 (2001).

\bibitem{Zhou-Kurths}
  C. Zhou and J. Kurths,
  Phys. Rev. Lett. {\bf 88}, 230602 (2002).

\bibitem{Pakdaman} K. Pakdaman, Neural Comput. {\bf 14}, 781 (2002).

\bibitem{Teramae-Dan}
  J. Teramae and D. Tanaka,
  Phys. Rev. Lett. {\bf 93}, 204103 (2004);
  Prog. Theoret. Phys. Suppl. {\bf 161}, 360 (2006).

\bibitem{Goldobin-Pikovsky}
  D. S. Goldobin and A. Pikovsky,
  Phys. Rev. E {\bf 71}, 045201(R) (2005);
  Physica A {\bf 351}, 126 (2005); Phys.  Rev. E {\bf 73}, 061906 (2006).

\bibitem{Nagai-Nakao}
  K. Nagai, H. Nakao, and Y. Tsubo,
  Phys. Rev. E {\bf 71}, 036217 (2005);
  H. Nakao, K. Nagai, and K. Arai, Prog. Theoret. Phys. Suppl. {\bf 161}, 294 (2006).

\bibitem{Nakao-Arai} 
  H. Nakao, K. Arai, K. Nagai, Y. Tsubo, and Y. Kuramoto, 
  Phys. Rev. E {\bf 72}, 026220 (2005).

\bibitem{Nakao-Arai-Kawamura}
  H. Nakao, K. Arai and Y. Kawamura,
  Phys. Rev. Lett. {\bf 98}, 184101 (2007).
  
\bibitem{Galan2}
  R. F. Gal\'an, G. B. Ermentrout, and N. N. Urban, Phys. Rev. E {\bf 76}, 056110 (2007).

\bibitem{Winfree}
  A. T. Winfree,
  {\it The Geometry of Biological Time} (Springer-Verlag, New York, 2001).

\bibitem{Kuramoto}
  Y. Kuramoto,
  {\it Chemical Oscillation, Waves, and Turbulence} (Springer-Verlag, Tokyo, 1984) (republished by Dover, New York, 2003).

\bibitem{Marcus} S. I. Marcus, IEEE Transactions on Information Theory
  {\bf IT-24}, 164 (1978).

\bibitem{Gray} R. A. Gray and N. Chattipakorn, Proc. Natl. Acad. Sci
  {\bf 102}, 4672 (2005).
  
\bibitem{Galan-Ermentrout-Urban} R. F. Gal\'an, G. B. Ermentrout and
  N. N. Urban, Phys. Rev. Lett. {\bf 94}, 158101 (2005).
  
\bibitem{Tateno-Robinson} T. Tateno and H. P. C. Robinson, Biophysical
  Journal {\bf 92}, 683 (2007).
  
\bibitem{Hanson}
  F. B. Hanson,
  {\it Applied Stochastic Processes and Control for Jump-Diffusions} (SIAM Books, 2007).

\bibitem{Mackey} A. Lasota and M. C. Mackey, {\it Probabilistic
    properties of deterministic systems} (Cambridge University Press,
  Cambridge, 1985).
 
\bibitem{Ott} E. Ott, {\it Chaos in Dynamical Systems} (Cambridge
  University Press, Cambridge, 2002).

\bibitem{Ermentrout-Saunders} G. B. Ermentrout and D. Saunders,
  J. Comput. Neurosci. {\bf 20}, 179 (2006).
  
\bibitem{Koch} C. Koch, {\it Biophysics of Computation} (Oxford
  University Press, Oxford, 1999).

\bibitem{Fujisaka-Yamada} H. Fujisaka and T. Yamada, Prog. Theor.
  Phys. {\bf 69}, 32 (1983); H. Fujisaka, Prog. Theor. Phys. {\bf
    70}, 1264 (1983); H. Fujisaka and T. Yamada, Prog. of Theor.
  Phys. {\bf 74}, 918 (1985).

\bibitem{Pikovsky} A. S. Pikovsky, Phys. Lett. A {\bf 165}, 33 (1992).

\bibitem{NakaoRMP}
  H. Nakao, Phys. Rev. E {\bf 58}, 1591 (1998).

\bibitem{Kitada} S. Kitada, Physica A {\bf 370} 539, (2006).

\bibitem{Pfeuty} B. Pfeuty, G. Mato, D. Golomb, and D. Hansel, 
Neural Comput. {\bf 17}, 633 (2005).

\bibitem{KawamuraCollective} Y. Kawamura, H. Nakao, K. Arai, H. Kori,
  and Y. Kuramoto,  Phys. Rev.  Lett. {\bf 101}, 024101 (2008).
\end{thebibliography}
\end{document}